\documentclass[aps,preprint,superscriptaddress,floatfix]{revtex4-1}
\usepackage{subfiles}
\usepackage{graphicx}
\usepackage{color}
\usepackage{amsmath}
\usepackage{float}
\usepackage{siunitx}
\usepackage{hyperref}
\usepackage{caption}
\usepackage{subcaption}
\usepackage[export]{adjustbox}
\usepackage[capitalise]{cleveref}
\usepackage{subcaption}

\hypersetup{
colorlinks=true,
urlcolor=blue,
citecolor=blue,
linkcolor=black
}

\begin{document}
\graphicspath{{Figures/}{../Figures/}}

\title{Supersymmetry, half--bound states, and grazing incidence reflection}
\author{D. A. Patient}
\email[]{dp348@exeter.ac.uk}
\affiliation
{
School of Physics and Astronomy, University of Exeter, EX4 4QL, Exeter, United Kingdom
}
\author{S. A. R. Horsley}
\affiliation
{
School of Physics and Astronomy, University of Exeter, EX4 4QL, Exeter, United Kingdom
}
\date{\today}
\begin{abstract}
Electromagnetic waves at grazing incidence onto a planar medium are analogous to zero energy quantum particles incident onto a potential well.  In this limit waves are typically completely reflected.  Here we explore dielectric profiles supporting optical analogues of `half--bound states', allowing for zero reflection at grazing incidence.  To obtain these profiles we use two different theoretical approaches: supersymmetric quantum mechanics, and direct inversion of the Helmholtz equation.
\end{abstract}
\maketitle

\section{Introduction}
    \label{sec:Intro}
    At grazing incidence, a wave is nearly always completely reflected from a surface.  The effect can be observed optically for almost any surface viewed at a shallow angle, where it behaves as a mirror.  This behaviour has been known for a long time, and is used to enhance quantum reflection~\cite{oberst2005}, and X-ray scattering (where the refractive index typically differs only slightly from unity)~\cite{compton1923}.  Conversely, it presents a problem for radar absorbers~\cite{dewitt1988}, and perfectly matched layers in numerical simulations~\cite{collino1998}.  In this work we investigate the problem of designing graded dielectric materials that do not reflect at grazing incidence.

\begin{figure}[h!]
    \begin{subfigure}[b]{.49\textwidth}
        \centering
        \includegraphics[width=\textwidth]{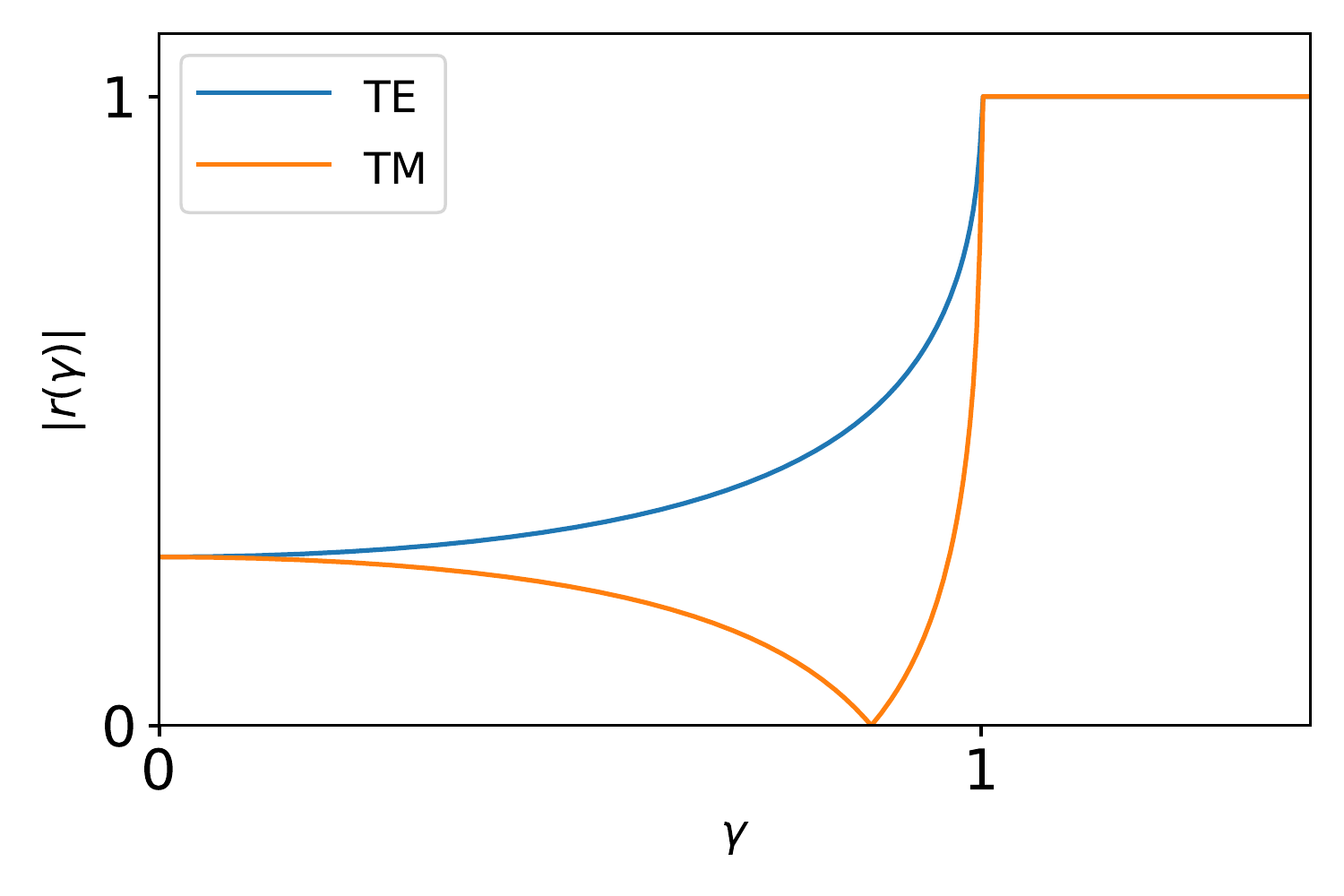}  
        \caption{Fresnel coefficients between two dielectric half spaces.\label{fig:Fresnel}}
    \end{subfigure}
    \begin{subfigure}[b]{.49\textwidth}
        \centering
        \includegraphics[width=\textwidth]{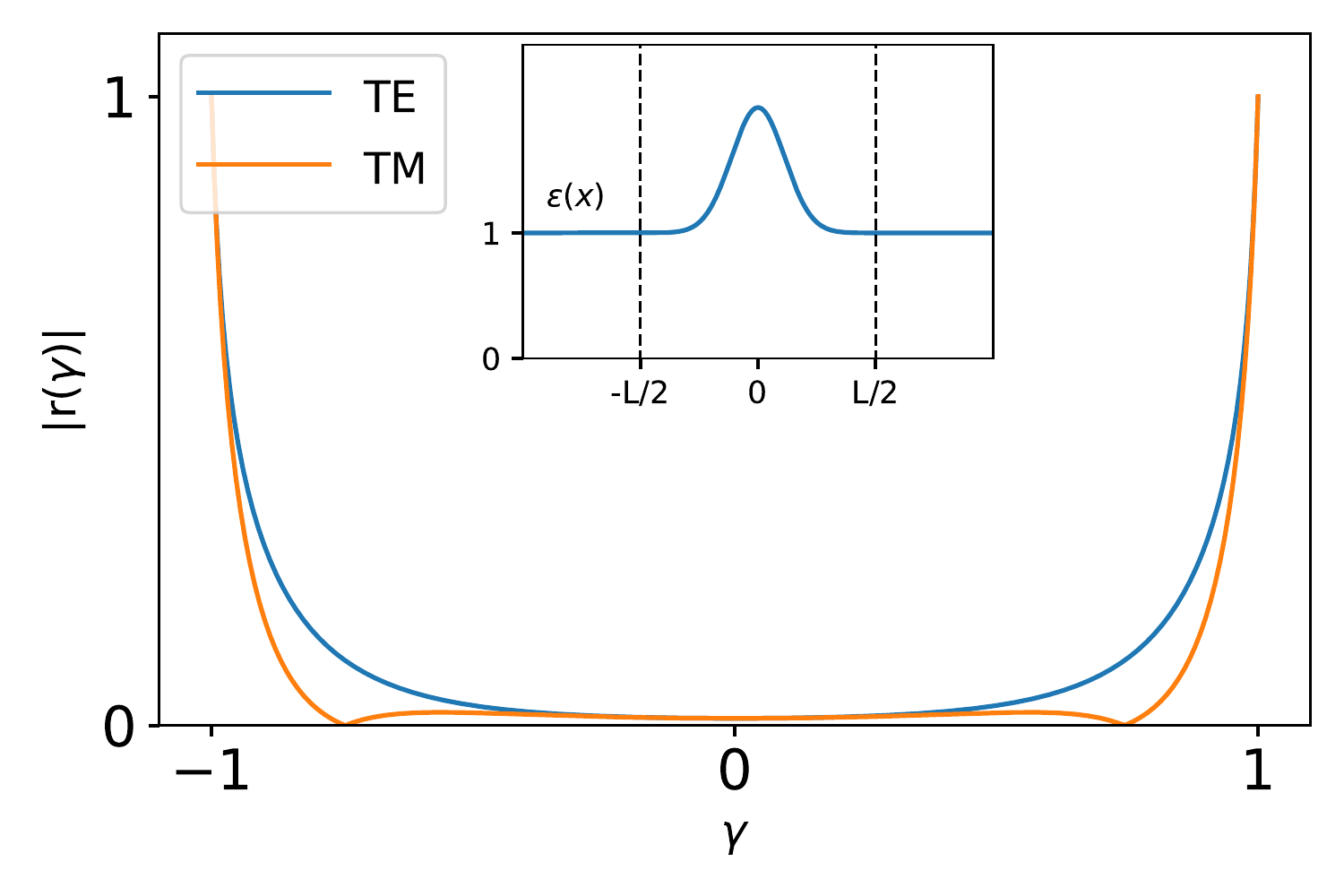}  
        \caption{Reflection coefficients for arbitrary permittivity profile (inset).\label{fig:Random}}
    \end{subfigure}
    \caption{Approaching grazing incidence ($\gamma = k_y/\sqrt{\epsilon_b}k_0 = 1$) the reflectivity of both TE and TM waves approaches unity.}
    \label{fig:Examples}
\end{figure}

Mathematically the phenomenon can be seen from a straightforward examination of the Helmholtz equation.  For an electromagnetic (EM) wave incident at an angle $\cos(\theta)=k_x/\sqrt{\epsilon_b}k_0$ onto a graded dielectric profile $\epsilon(x)$, and in terms of the dimensionless coordinate $\xi=k_x x$, the Helmholtz equation is
\begin{equation}
    \label{eq:TEEQ}
    \left[\frac{d^{2}}{d \xi^{2}}+1+\left(\frac{k_0}{k_x}\right)^2(\epsilon(\xi)-\epsilon_b)\right]\phi(\xi)=0
\end{equation}
where $\epsilon_b$ is the background value of the permittivity, such that $\epsilon(\xi)\to\epsilon_b$ as $|\xi|\to\infty$.  These coordinates are scaled such that away from the inhomogeneity the wave takes the form ${\rm e}^{\pm{\rm i}\xi}$ irrespective of the angle of incidence.  The effect of the angle is now subsumed in an effective permittivity profile $(k_0/k_x)^2\epsilon(\xi)$. This effective permittivity becomes infinitely large at grazing incidence, when $k_x\to0$.  A general permittivity profile will therefore act as a perfect reflector in this limit.  As a concrete example, take the Fresnel coefficients for the transverse electric (TE) and transverse magnetic (TM) polarizations~\cite{LandauEM1960}
\begin{equation}
    \label{eq:Fresnel}
    \begin{split}
        r_{TE} = \frac{\mu_2 k_{x,1} - \mu_1 k_{x,2}}{\mu_2 k_{x,1} + \mu_1 k_{x,2}} \\
        r_{TM} = \frac{\epsilon_2 k_{x,1} - \epsilon_1 k_{x,2}}{\epsilon_2 k_{x,1} + \epsilon_1 k_{x,2}},
    \end{split}
\end{equation}
where the normal components of the wave--vectors are $k_{x,n}=\sqrt{\epsilon_n \mu_n k_0^2 - k_y^2}$.  At grazing incidence $k_{x,1}\to0$, it is clear from (\ref{eq:Fresnel}) that both TE and TM reflectivities tend to unity, as shown in Fig.~\ref{fig:Fresnel}.  In Fig.~\ref{fig:Random} we show that this is also the case for an arbitrarily chosen dielectric profile (here a Gaussian variation of $\epsilon(x)$). 
%
%
Despite a low reflectivity of this smooth profile at normal incidence, the reflection nevertheless becomes complete at grazing incidence.  One can picture this as analogous to a stone skimming on the surface of water.  As the stone's momentum becomes close to parallel to the surface, only a small impulse is required to reverse the stone's motion normal to the surface.

Despite this general behaviour, surprisingly there are permittivity profiles (and analogous graded sound speeds in acoustics, or potentials $V(x)$ in quantum mechanics) that do \emph{not} act as perfect reflectors at grazing incidence.  The P\"{o}schl-Teller potential~\cite{Poschl1933} is perhaps the most famous example of such a non--reflecting profile: at some frequencies this potential does not reflect radiation at \emph{any} angle of incidence, despite a potentially rapid spatial variation of the profile.  This profile has been experimentally realised as a multilayer structure~\cite{thekkekara2014}, and its non--reflecting behaviour is understood in terms of inverse scattering theory and the solitons of the Kortweg--de Vries equation~\cite{gardner1967,drazin1989,horsley2016}.  Another example is provided by anisotropic magnetodielectric \emph{transformation media}~\cite{Pendry2000,leonhardt2006} (where $\boldsymbol{\epsilon}=\boldsymbol{\mu}$), which enact an effective transformation of the coordinate system.  Given this equivalence these materials cause no scattering whatsoever, for grazing incidence or otherwise.  Finally, isotropic dielectric materials obeying the spatial Kramers--Kronig relations~\cite{horsley2015} also do not reflect any wave incident from one side, although the limit of grazing incidence is particularly delicate for these profiles, requiring them to be infinitely extended~\cite{horsley2016b}.

Here we take a different approach. EM waves approaching an interface at grazing incidence are analogous to zero energy quantum particles incident onto a potential well.  It is known that there exist threshold anomalies \cite{Senn1988}, due to presence of so--called half-bound states (HBS) \cite{Senn1988}, allowing complete transmission through the potential well at zero energy~\cite{Wigner1948}.  The optical analogue of these anomalies therefore allow the complete transmission of grazing incidence waves.  The naming of these half-bound states originates in Levinson's theorem~\cite{levinson1949} (see~\cite{zhong2006} for a interesting review, and~\cite{wellner1964} for a simple proof), which connects the number of bound states of a potential to the phase shift of scattering in the zero energy limit.  As the name suggests, half-bound states count as half a bound state in the phase shift.  These states are non-normalizable `bound' states with zero energy.

To find materials supporting these half-bound states we take two theoretical approaches.  The first is to apply the factorisation method~\cite{Cooper1995} to the Helmholtz equation. This method is the equivalent of the factorization of a Hamiltonian into raising and lowering operators, and by requiring that the `lowering' operator has a zero eigenvalue we obtain a permittivity profile which doesn't reflect grazing incidence waves. This approach connects with recent work on analogues of supersymmetric quantum mechanics (SUSY) in optics~\cite{Chumakov1994,miri2013,longhi2013,yu2016,Miri2014,Meca2020}, where the same factorization is carried out to obtain isospectral structures.  The difference in the isospectral structures is in the removal of a single state.  In our particular case that single state is the half bound state, which allows for zero reflection at grazing incidence.

In addition to factorization of the Helmholtz equation, we derive another set of permittivity profiles from an inversion of the Helmholtz equation, where the permittivity is written in terms of the wave amplitude.  As we'll see, this yields similar results to the factorization approach, but with the freedom to specify additional boundary conditions.  As an example, we'll derive a graded abosrbing dielectric layer that can be added to a mirror, and which removes reflection at grazing incidence.
    
\section{Grazing Incidence Reflection and Half-Bound States}
    \label{sec:SUSYExplained}
    We first review the problem of vanishing reflection at grazing incidence, and it's relation to so--called `half-bound states'.  Consider a plane wave incident from the left of a dielectric profile $\epsilon(x)$, the profile becoming homogeneous as $|x|\to\infty$. For zero reflection, the wave must be of the form $\exp(i k_x x)$ on the far left and far right of the profile. Outside the slab the Helmholtz equation becomes, in the limit of grazing incidence, Laplace's equation $d^{2}\phi/dx^{2}=0$ with solutions $a+b x$, where $a$ and $b$ are constants.  But if there is to be zero reflection in this limit, outside the slab we must have $\exp(i k_x x)\to 1 + i k_x x$.  As $k_x\to 0$ the wave equals unity everywhere outside the slab.  Therefore if there is to be zero reflection in the limit of grazing incidence the Helmholtz equation must have two independent solutions, one which is equal to a constant on both sides of the slab, and the other proportional to the $x$ coordinate, with the same proportionality constant on both sides of the slab.

Provided the permittivity profile is symmetric about its centre (so that both $\phi(x)$ and $\phi(-x)$ are solutions), and that one of the solutions to the TE field is constant outside the slab, we thus ensure zero reflection at grazing incidence (although non--symmetric potentials can also exhibit less than complete reflection at grazing~\cite{nogami1996}).  In quantum mechanics a state which is constant outside of a potential where it is `bound' is not normalizable, and is known as a `half-bound state' (see e.g. page 280 of~\cite{newton2002} for a classification of the bound states of potentials).  These states are bound states with an energy that is at the edge of the continuum.  In the reflection from a dielectric profile we can understand these states as wave--guide modes whose decay constant is close to zero. From this point of view we are providing a method for designing waveguides with one mode with vanishing decay constant outside the guide.

\begin{figure}
    \centering
    \includegraphics[width=0.5\columnwidth,valign=m]{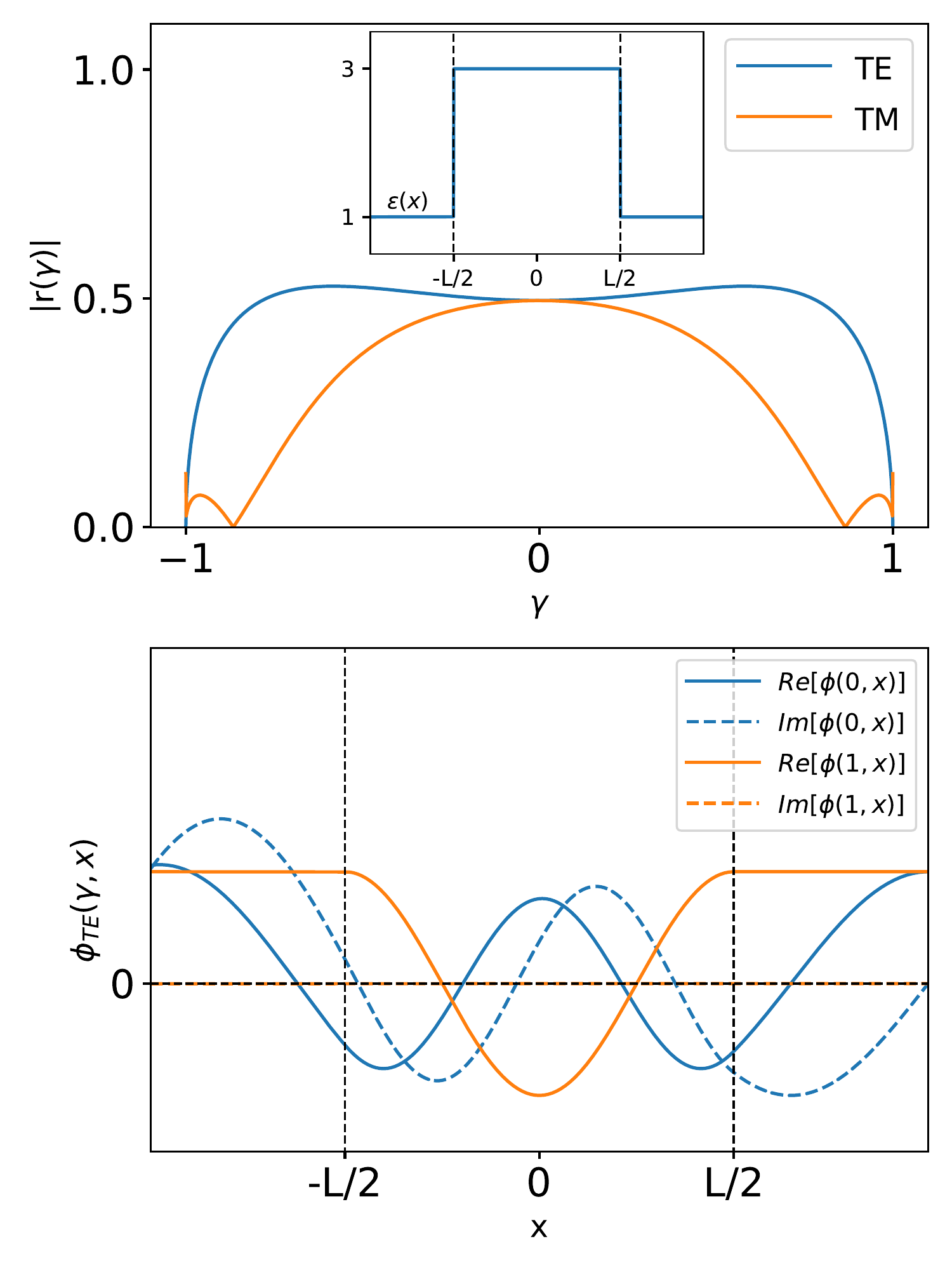}
    \caption{(top) Reflectivity as a function of angle $\gamma=\sin(\theta)$ for a homogeneous slab (permittivity profile inset) in free space ($\epsilon_b = 1$), whose length obeys \cref{eq:SlabLength} for $\epsilon = 3$, $k_0 = 1$ and $n = 2$. (bottom) the field of a TE wave indicent onto the slab (boundaries indicated as vertical dashed lines), in the cases where the initial wave is at normal incidence (blue) and close to grazing (orange). At grazing incidence, the field decays to a non-zero constant outside the region where the dielectric exists, indicating that the slab supports a HBS.}
    \label{fig:Homog}
\end{figure}

The simplest example of a system supporting a half-bound state is a homogeneous slab waveguide of width $L$, with a constant permittivity $\epsilon(x) = \epsilon$ that spans a region $x \in [-L/2,L/2]$.  Re--introducing the coordinate $x$,  (\ref{eq:TEEQ}) becomes
\begin{equation}
    \label{eq:TEHHDimensionless}
    \left[ \frac{d^2}{dx^2} + k_0^2\left( \epsilon(x) - \epsilon_b \right) + k_x^2 \right] \phi(x) = 0,
\end{equation}
The solutions within the two regions are given by $\phi(x)=\exp(\pm{\rm i}k_x x)$ ($|x|>L/2$) and $\phi(x)=\exp(\pm{\rm i}\sqrt{k_0^2(\epsilon-\epsilon_b) + k_x^2}x)$ ($|x|<L/2$).  If the phase accumulated inside the slab is an integer number of $\pi$, the field on the right of the slab is $\pm$ the field on the left, and the boundary conditions can be fulfilled without any reflected wave.  This condition is given by
\begin{equation}
    \label{eq:HBSLimit}
    \exp \left[\pm {\rm i} \sqrt{k_0^2(\epsilon - \epsilon_b) + k_x^2}L\right] = \pm 1 = \exp({\rm i} n \pi),
\end{equation}
a condition which can be fulfilled with a slab of length $L=n\pi/\sqrt{k_0^2(\epsilon - \epsilon_b) + k_x^2}$, which is the standard condition for a transmission resonance.  If we choose the angle of zero reflection to be grazing incidence, then the slab waveguide must have width 
\begin{equation}
    L=\frac{n\pi}{k_0\sqrt{\epsilon-\epsilon_b}}\label{eq:SlabLength}
\end{equation}
In this case the field outside the slab reduces to a constant at grazing incidence, which is the half-bound state just mentioned.  \cref{fig:Homog} shows the numerically computed reflectivity as a function of angle for a homogenous slab of permittivity $\epsilon = 3$ and background permittivity $\epsilon_b =1$ whose length satisfies \cref{eq:SlabLength}.  For the numerical integration we apply the odeint function from the SciPy library~\cite{2020SciPy-NMeth} to integrate Eq.~(\ref{eq:TEHHDimensionless}).  As the figure shows, the TE reflectivity is zero as we approach $|\gamma|=|\sin(\theta)|=1$.
\subsection{Factorisation Method}
    \label{sec:Factorise}
    In this section, we use the factorisation method usually applied in supersymmetric quantum mechanics~\cite{Cooper1995} to design a family of HBS supporting dielectric profiles, which do not reflect TE waves at grazing incidence.  We note that our approach connects with recent work in quantum mechanics by Ahmed et al. where zero energy reflectionless potentials have been investigated in several cases~\cite{ahmed2017b,dijk2017,ahmed2020}, and where half-bound state supporting potentials have been subject to supersymmetric transformations to generate e.g. potentials supporting no bound states at all~\cite{ahmed2017}.
    
    Here we start by writing the Helmholtz equation (\ref{eq:TEHHDimensionless}) as a product of first order operators
    \begin{equation}
        \label{eq:HHOperators}
        \left( - \frac{d}{dx} + k_0 \alpha(x) \right) \left( \frac{d}{dx} + k_0 \alpha(x) \right) \phi(x) = \hat{a}^{\dagger} \hat{a} \phi(x) = 0,
    \end{equation}
    where $\hat{a}^{\dagger}, \hat{a}$ are analogous to raising and lowering operators in quantum mechanics (although their commutator is not proportional to a constant).  This factorization can only be carried out for certain forms of the permittivity $\epsilon(x)$.  Assuming grazing incidence $k_x=0$, expanding out the product in (\ref{eq:HHOperators}), and comparing to \cref{eq:TEHHDimensionless} gives the relationship between the function $\alpha(x)$, and the material parameter $\epsilon(x)$
    \begin{equation}
        \label{eq:Alpha}
        \epsilon(x)-\epsilon_b = \frac{1}{k_0} \frac{d \alpha(x)}{dx} - \alpha^2(x).
    \end{equation}
    Having factorized the Helmholtz equation in the form (\ref{eq:HHOperators}) we can immediately find the solution where $\hat{a}\phi(x)=0$
    \begin{equation}
        \label{eq:FieldSol1}
        \phi_1(x) = \phi_1(-L/2) \exp \left[ -k_0 \int_{-L/2}^x \alpha(x') dx' \right].
    \end{equation}
    As described earlier, the second solution does not need to be considered. By inspection of \cref{eq:Alpha}, one can see that if $\alpha(x) \rightarrow 0$ as $|x| \rightarrow \infty$, then $\epsilon(x) = \epsilon_b$ as required. For vanishing reflection of grazing incidence waves, the amplitude of the wave outside of the slab must be the same on either side. This requires $\int_{-L/2}^{L/2} \alpha(x) dx = 0$, a condition which we enforce by choosing $\alpha(x)$ antisymmetric around $x=0$.  Unlike the case of a true bound state where, for a given $\alpha(x)$, one of the kernels of $\hat{a}$ and $\hat{a}^{\dagger}$ will diverge at infinity, here the functions satisfying $\hat{a}\phi=0$ and $\hat{a}^{\dagger}\phi=0$ \emph{both} represent half bound states.  This means that the operators can be ordered either way in (\ref{eq:HHOperators}) and both corresponding profiles will support HBS and have vanishing reflection at grazing incidence.  To be precise, both the profile given in Eq. (\ref{eq:Alpha}) and that where $\epsilon(x)-\epsilon_b=-k_0^{-1}d\alpha/dx-\alpha^{2}(x)$ have zero reflection at grazing incidence.
    
\begin{figure}
        \centering
        \includegraphics[width=0.5\linewidth,valign=m]{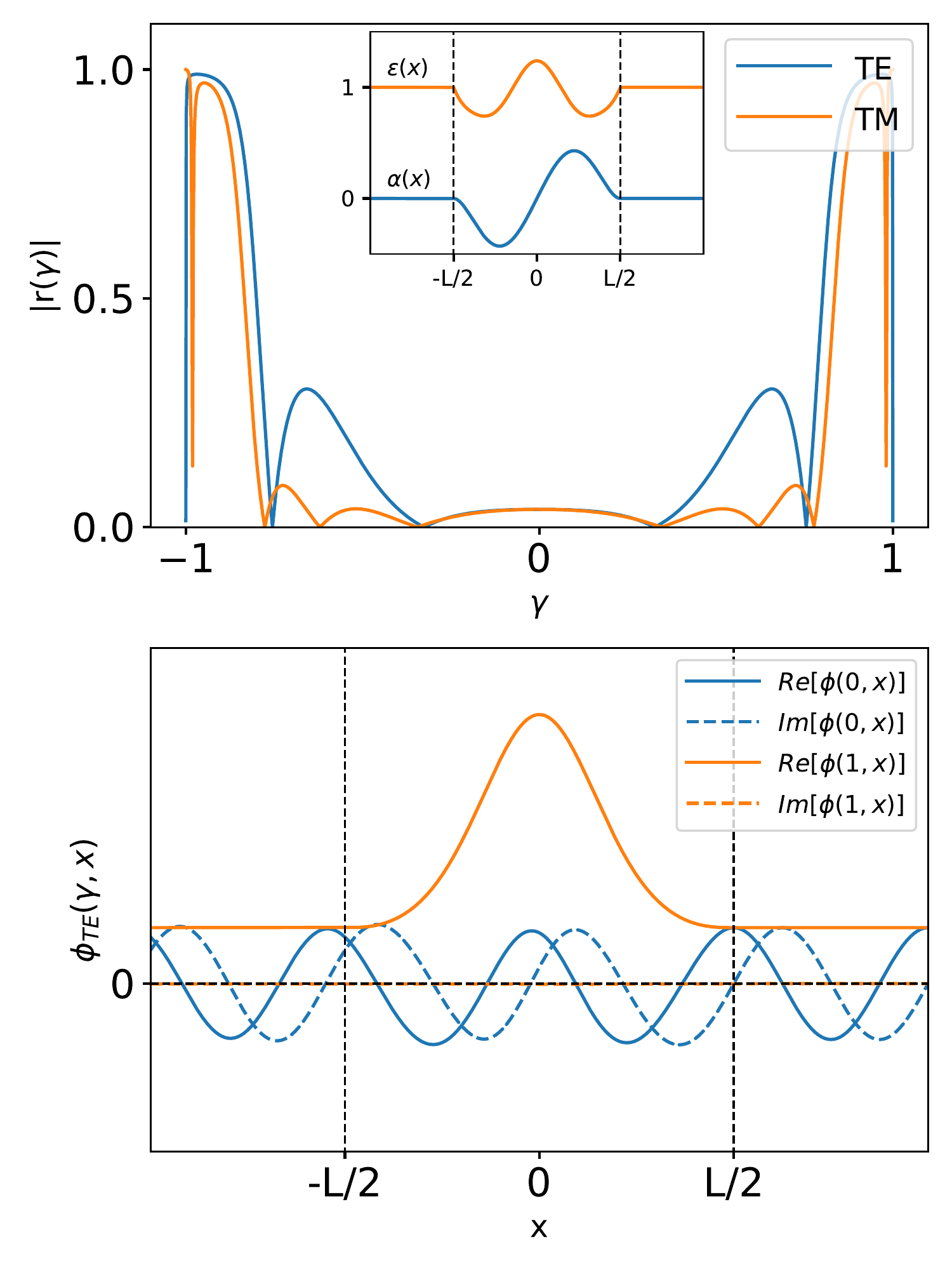}
        \caption{Reflectivity as a function of angle $\gamma=\sin(\theta)$ (top) for a permittivity ($\epsilon_b = 1$) generated using (\ref{eq:Alpha}) with both permittivity and $\alpha(x)$ inset.  In this example there are large regions where $\epsilon(x) < 1$ which would be expected to cause extremely strong reflection as the angle of incidence increases.  The onset of this strong reflection is evident beyond $\gamma\sim0.8$ after which the reflectivity approaches unity.  Rather surprisingly as the angle becomes close to grazing, this strong reflection rapidly decreases to zero.  Note that a dip in reflectivity as we approach grazing incidence is evident for both polarizations.  The TE field (bottom) is shown in the case of normal (blue) and grazing incidence (orange), demonstrating that---in accordance with our design---at grazing incidence there is a mode that is constant outside the slab.}
        \label{fig:SUSY1}
    \end{figure}

\begin{figure}
        \centering
        \includegraphics[width=0.5\linewidth,valign=m]{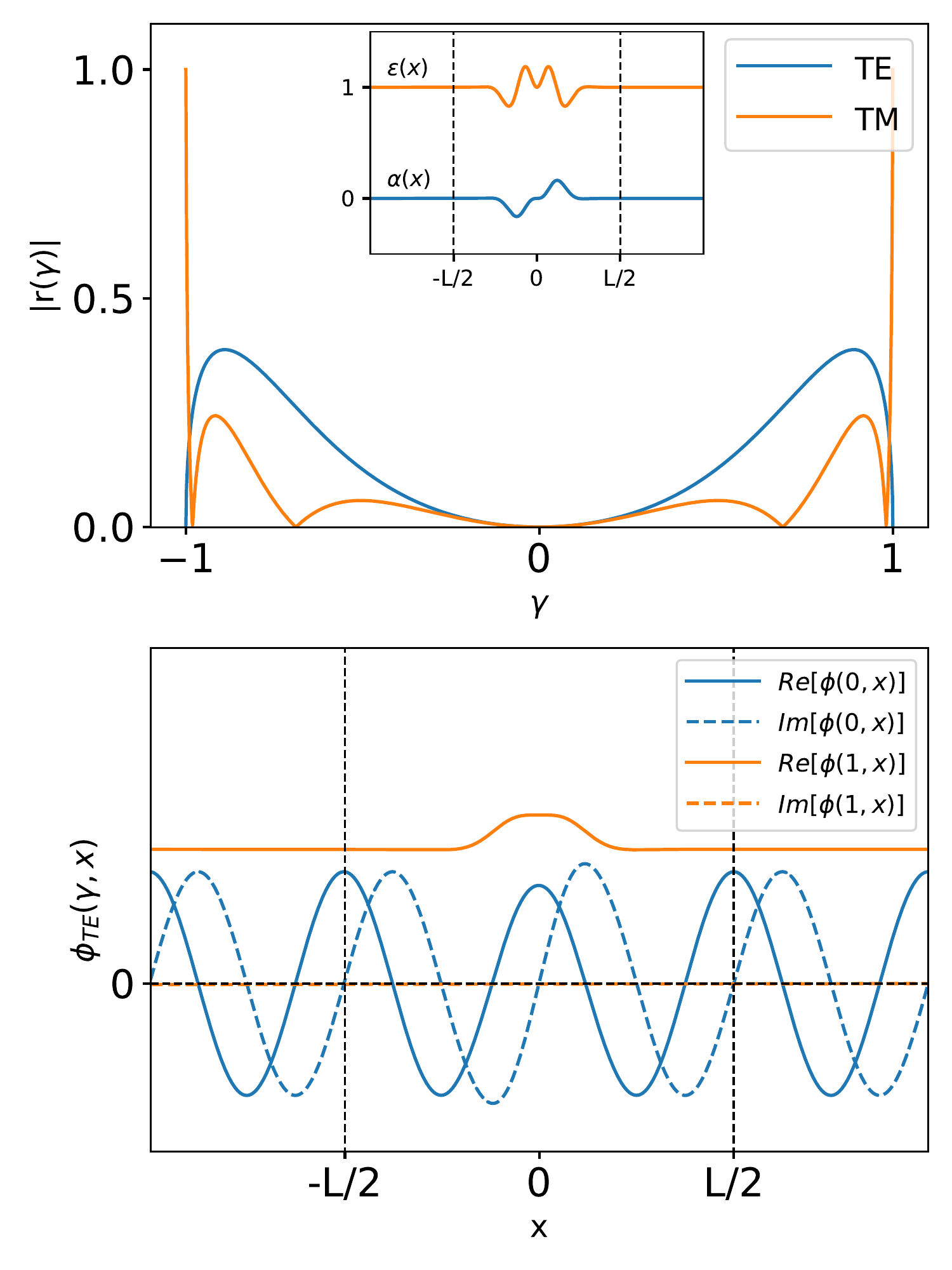}
        \caption{As in \cref{fig:SUSY1}, for a different choice of $\alpha(x)$ profile. In this case, $\alpha(x)$ is chosen such that the region where $\epsilon(x) < 1$ is reduced. Accordingly the reflectivity reduces to zero over a greater angular range than in \cref{fig:SUSY1}, as we approach grazing incidence.  Again notice that the TM polarization also exhibits a dip in reflectivity close to $|\gamma|=1$.}
        \label{fig:SUSY2}
    \end{figure}

    With these conditions, an entire family of these profiles can be generated. Two such examples are given in \cref{fig:SUSY1,fig:SUSY2}. To generate these HBS supporting profiles, we can choose any odd function $\alpha(x)$, with a value and derivative that goes to zero at $|x|=L/2$. The permittivity profiles derived from \cref{eq:Alpha} then give us designs for dielectric materials that don't reflect TE waves at grazing incidence. Interestingly, with $\alpha(x)$ being an odd function, the permittivity profiles necessarily have regions where $\epsilon(x) < \epsilon_b$.  This is immediately evident from (\ref{eq:Alpha}): after increasing in magnitude away from $x=-L/2$, $\alpha(x)$ must return to zero at $x=+L/2$.  Therefore at some point the function $\alpha(x)$ must turn around, having finite value but zero gradient.  At this point $\epsilon(x)-\epsilon_b=-\alpha^2<0$.  Close to grazing the wavevector in the direction of propagation, $k_x$ becomes complex.  In e.g. the WKB method, at the turning point where $\epsilon=\epsilon_b$, the wave is matched to a combination of Airy functions.  In the case of a single turning point this gives rise to complete reflection~\cite{heading2013}.  The profiles found here have several turning points and run counter to this intuition, allowing complete transmission, despite the regions of exponential decay.  Nevertheless the intuition is not completely wrong.  The larger the region where $\epsilon(x)$ dips below $\epsilon_b$, the larger the reflection is \emph{close} to grazing incidence, and the narrower the range of angles close to $|\sin(\theta)|$ where the reflection vanishes.  This effect can be seen in \cref{fig:SUSY1}, where the comparatively large region of  $\epsilon(x) < \epsilon_b$, leads to a mirror-like behaviour above $\gamma\sim0.8$, and then a very narrow angular range of low reflectivity around grazing incidence. Conversely in \cref{fig:SUSY2} the region where $\epsilon(x) < \epsilon_b$ is relatively small, resulting in a larger region of low reflectivity close to grazing incidence.  It is an interesting question whether those profiles with a very sharp drop in reflectivity close to grazing could be realised with sufficient precision (using e.g. dielectric multilayers or metamaterials) to make this narrow angular region of low reflectivity observable.
    
\section{Inversion of the Helmholtz Equation}
    \label{sec:Phase}
    The factorization method is not the only way to find profiles that don't reflect grazing incidence waves. We now show that such designs can also be found via a direct inversion of the Helmholtz equation. This, for example allows us to have find complex permittivity profiles, and to modify the system's boundary conditions.

\begin{figure}
        \centering
        \includegraphics[width=0.5\linewidth,valign=m]{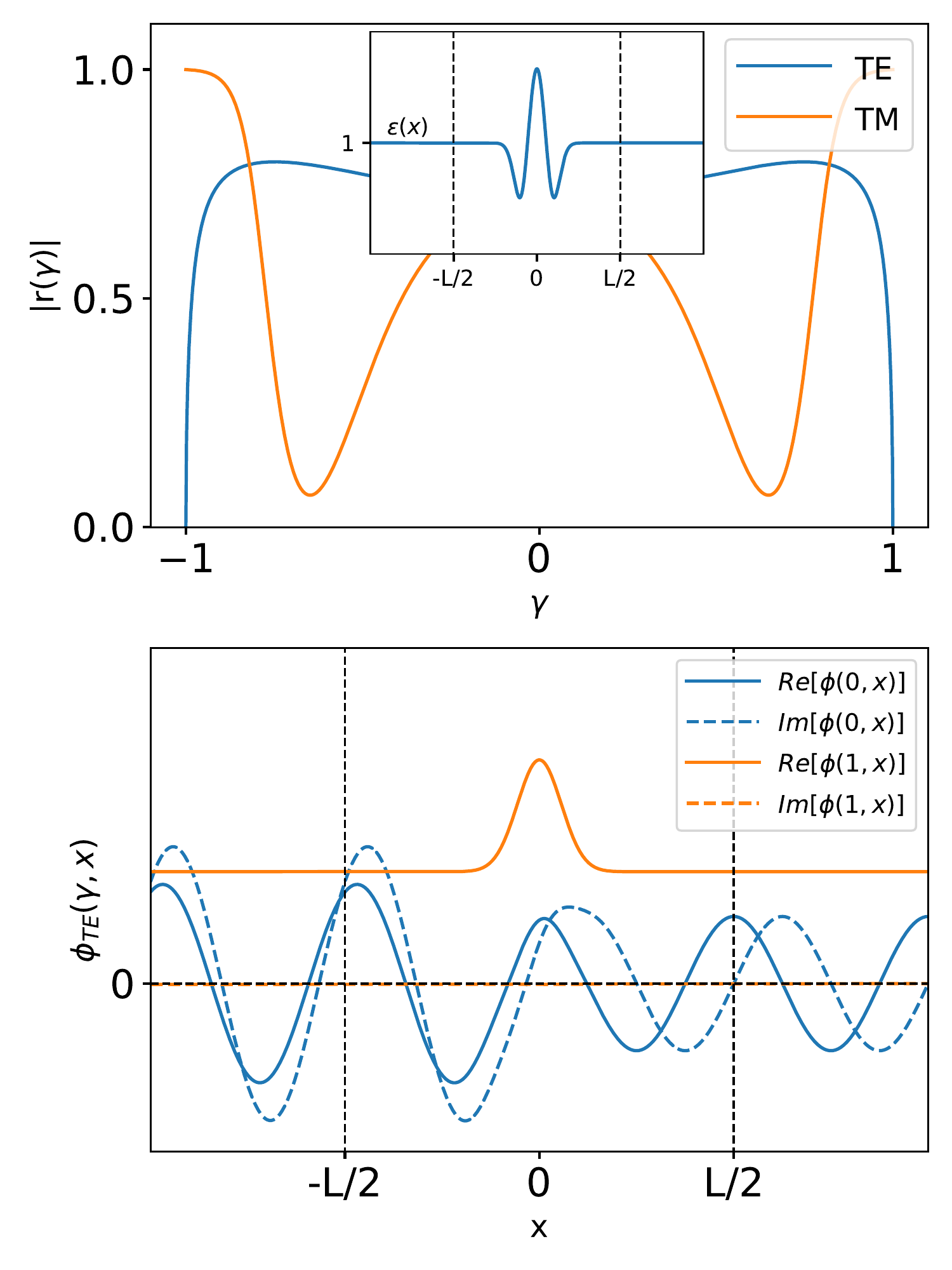}
        \caption{Reflectivity as a function of angle (top) for the permittivity profile (\ref{eq:PermForHBS}) (inset) found through inversion of the Helmholtz Equation at grazing incidence. The field (bottom) of the waves approaching at normal (blue) and grazing incidence (orange) show that as the angle approaches grazing incidence, the wave approaches the half-bound state (\ref{eq:HBSField}).}
        \label{fig:HBSField}
    \end{figure}

Take \cref{eq:TEHHDimensionless} with a background permittivity $\epsilon_b = 1$. Inverting the equation such that the permittivity is given as a function of the field, we obtain the simple equation
\begin{equation}
    \label{eq:InverseHHSimple}
    \epsilon(x) = 1 - \left(\frac{k_x}{k_0}\right)^2- \frac{\phi''(x)}{k_0^2\phi(x)},
\end{equation}
where $\phi''(x)$ denotes the second derivative of the field w.r.t position. From our previous discussion we know that if, at grazing incidence the field is constant outside a profile which is symmetric in space, then the reflection vanishes.  As a first example we assume the field profile is a Gaussian centred at $x=b$ with width $a$, plus a constant $A$
    \begin{equation}
        \label{eq:HBSField}
        \psi(x) = A + \exp\left[ -\left( \frac{x - b}{a} \right)^2 \right].
    \end{equation}
    Provided $A \neq 0$, this function is a half-bound state. Inserting this into \cref{eq:InverseHHSimple} with $k_x=0$ we find the permittivity profile
    \begin{equation}
        \label{eq:PermForHBS}
        \epsilon(x) = 1 - \left[ \frac{4(x - b)^2 - 2a^2}{a^4 k_0^2 \left(A \exp \left[ \left( \frac{x - b}{a} \right)^2 \right] + 1\right)} \right].
    \end{equation}
    A numerical calculation of the reflectivity from this permittivity profile is given in \cref{fig:HBSField}. The TE reflectivity decays to zero at grazing incidence. There are again regions in the permittivity profile where $\epsilon(x) < 1$. These occur on the outer portions of the Gaussian where the second derivative $\phi''$ is positive.  Again we find profiles with regions where a wave at grazing incidence decays exponentially, but yet does not reflect.  It is not however necessary for the permittivity to have such regions: it is possible to have $\phi''(x)/\phi(x)<0$ everywhere, with zero gradient and non--zero value at $|x|=L/2$, as in the case of a homogeneous slab of constant permittivity where $\phi=\cos(k_0\sqrt{\epsilon-1}x)$.
    
    While it is interesting that there are a plethora of dielectric materials that do not reflect grazing incidence waves, in practice it would be useful to reduce grazing incidence reflection from an otherwise reflecting object.  As an example, we consider applying the above theory to design an absorbing layer that when placed onto a mirror removes grazing incidence reflection at a fixed frequency $f=c k_0/2\pi$.  We assume the following ansatz for the field
    \begin{equation}
        \label{eq:FieldWithMirror}
        \phi(x) =   \begin{cases}
                        e^{i k_x x} & {x <0} \\
                        a (x -  L) + b (x -  L)^3 & {0 \leq x \leq L} \\
                        0 & {x > L},
                    \end{cases}
    \end{equation}
    i.e. an incident wave at angle $k_x/k_0=\cos(\theta)$ is completely absorbed in a layer of width $L$. The choice of the field inside the layer is a chosen such that the term $\phi''(x)/\phi(x)$ does not diverge at any $x$, and such that the field is zero on the mirror at $x = L$.  Continuity of the field and its first derivative at $x=0$ gives the two constants $a$ and $b$
    \begin{equation}
        \label{eq:CoeffsNoReflectMirror}
        \begin{split}
            a = \frac{- 3 - {\rm i} k_x L}{2 L} \\
            b = \frac{1 + {\rm i} k_x L}{2 L^3}.
        \end{split}
    \end{equation}
    Inserting \cref{eq:FieldWithMirror} into \cref{eq:InverseHHSimple} then yields, in the slab region
    \begin{equation}
        \label{eq:MirrorEpsRequired}
        \epsilon(x) = 1 - \frac{6}{k_0^{2}}\frac{1}{(\frac{a}{b}) + (x -  L)^2}=1+\frac{3}{(k_0 L)^{2}}\frac{1}{(\frac{1-{\rm i}k_x L}{1+k_x^{2} L^{2}})+\frac{1}{2}\left[1-\left(\frac{x}{L} -  1\right)^2\right]},
    \end{equation}
    with $\epsilon(x)=1$ where $x<0$.  This complex profile removes reflection at a given angle $\theta$, determined by $k_x/k_0=\cos(\theta)$.  The imaginary part of the profile represents the absorption required to eliminate the wave before it reaches the mirror, and is positive throughout the profile.
        
    \begin{figure}[h!]
        \centering
        \includegraphics[width=0.7\columnwidth]{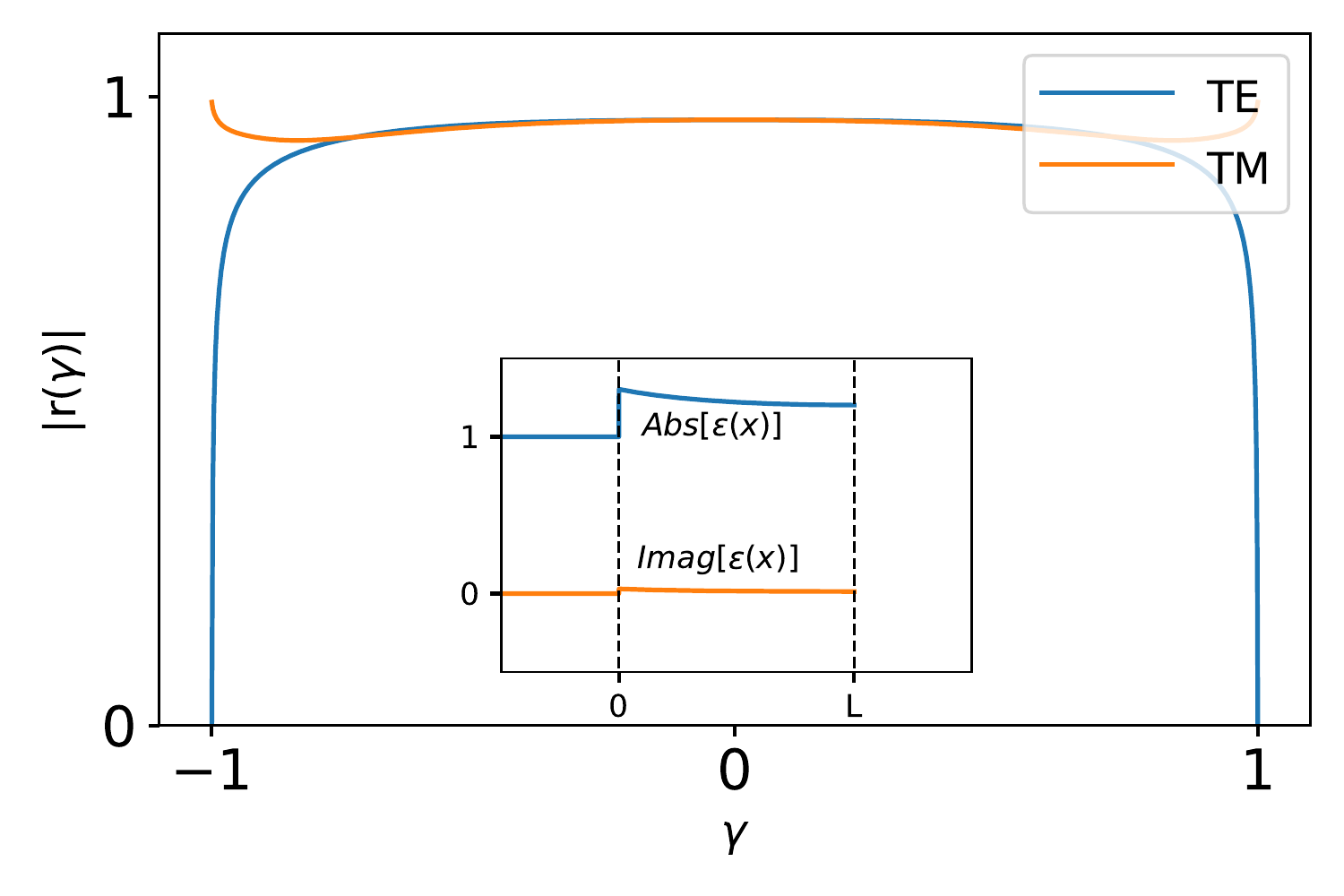}
        \caption{The geometry of the system is such that waves are incident from the left onto a dielectric material that is flanked by a perfect mirror. The reflectivity as a function of angle for the chosen permittivity function (\cref{eq:MirrorEpsRequired}) is designed such that the reflection coefficient close to grazing incidence is zero.}
        \label{fig:NoReflectWithMirror}
    \end{figure}
    
    \cref{fig:NoReflectWithMirror} shows the reflectivity as a function of angle for the permittivity function \cref{eq:MirrorEpsRequired} for the case of grazing incidence $k_x \approx 0$.  Interestingly, in this limit the loss in the system becomes negligible and the profile tends to the limiting form $\epsilon(x)=1+6(k_0 L)^{-2}[1-(x/L-1)^2]^{-1}$.  This is paradoxical, as---due to energy conservation---away from grazing incidence the addition of a lossless layer to a mirror cannot give anything other than a phase shift to the reflected wave.  The resolution to this paradox is that at grazing incidence the component of the Poynting vector normal to the mirror is zero, so we are not violating energy conservation.  As the angle of zero reflection is brought towards grazing incidence, the loss in the profile diminishes.  When the profile is to be reflectionless at exactly grazing incidence, we have a profile that has a reflectivity of unity except at exactly $k_x=0$ where the dielectric acts as a waveguide with a confined mode that has an infinite decay constant in the region of free space outside.  When $k_x$ is close to zero the profile has a small amount of loss, which serves to absorb the wave and eliminate reflection close to grazing incidence.  This shows that only a small amount of loss is required to eliminate the reflection of a wave where $k_x\sim0$.

\section{Conclusions}
    \label{sec:Conclude}
    By factorizing the Helmholtz equation into a product of operators (as is done in supersymmetric quantum mechanics) we were able to design graded dielectric profiles that do not reflect TE polarized electromagnetic waves at grazing incidence.  The physics behind this absence of reflection is analogous to the `threshold' anomalies previously identified in one--dimensional quantum scattering, and occurs due to the presence of a zero `energy' bound state within the profile.  In electromagnetism we can understand these as waveguide modes that have an infinite decay constant, and our design procedure produces graded index waveguides with one mode that is on the boundary of being confined or radiative.  This illustrates a new application of the formalism of supersymmetric quantum mechanics to the design of electromagnetic materials.  The same approach could equally be applied to the TM polarization, if $\mu(x)$ was graded instead of the permittivity $\epsilon(x)$.  We found that the graded profiles derived in this way require regions where the permittivity is less than the background level $\epsilon_b$.  This is counter--intuitive, as such values usually imply strong reflection, leading to total internal reflection in the limit of an infinitely thick homogeneous slab.

In addition we also inverted the Helmholtz equation to obtain the permittivity in terms of the form of the wave at grazing incidence.  From this we were again able to find similar profiles exhibiting a half-bound state as the incident wave approaches grazing.  The inversion of the Helmholtz equation has the advantage that boundary conditions can be imposed on the wave.  Imposing the vanishing of the wave amplitude at a single point, we found an absorbing graded index profile that can be added to the surface of a mirror to remove reflection at a particular angle.  As the zero-reflection angle approaches grazing incidence the dissipation in the absorbing layer tends to zero, and there is an interesting regime where the reflection close to grazing is very small as is the dissipation in the layer.
    
\begin{acknowledgments}
    The authors would like to acknowledge the Exeter Metamaterials CDT and the EPSRC (EP/L015331/1) for funding and supporting this research.  SARH acknowledges financial support from a Royal Society TATA University Research Fellowship (RPG-2016-186).
\end{acknowledgments}

\bibliography{main}

\end{document}